\documentclass[twocolumn,preprintnumbers,amsmath,amssymb,superscriptaddress,pre,showpacs]{revtex4}
\usepackage{graphicx}% Include figure files
\usepackage{dcolumn}% Align table columns on decimal point
\usepackage{bm}% bold math
\usepackage[utf8]{inputenc}

\def\acc{$\mathrm{m\,s^{-2}}$}

\begin{document}

\title{Granular convection and the Brazil nut effect in reduced gravity}

\author{Carsten Güttler}
\affiliation{Department of Earth and Planetary Sciences, Kobe University, 1-1 Rokkodai-cho, Nada-ku, Kobe 657-8501, Japan}
\affiliation{Institut für Geophysik und extraterrestrische Physik, Technische Universität Braunschweig, Mendelssohnstraße 3, D-38106 Braunschweig, Germany}
\author{Ingo von Borstel}
\affiliation{Institut für Geophysik und extraterrestrische Physik, Technische Universität Braunschweig, Mendelssohnstraße 3, D-38106 Braunschweig, Germany}
\author{Rainer Schräpler}
\affiliation{Institut für Geophysik und extraterrestrische Physik, Technische Universität Braunschweig, Mendelssohnstraße 3, D-38106 Braunschweig, Germany}
\author{Jürgen Blum}
\affiliation{Institut für Geophysik und extraterrestrische Physik, Technische Universität Braunschweig, Mendelssohnstraße 3, D-38106 Braunschweig, Germany}

\date{\today}

\begin{abstract}
We present laboratory experiments of a vertically vibrated granular medium consisting of 1 mm diameter glass beads with embedded 8 mm diameter intruder glass beads. The experiments were performed in the laboratory as well as in a parabolic flight under reduced-gravity conditions (on Martian and Lunar gravity levels). We measured the mean rise velocity of the large glass beads and present its dependence on the fill height of the sample containers, the excitation acceleration, and the ambient gravity level. We find that the rise velocity scales in the same manner for all three gravity regimes and roughly linearly with gravity.
\end{abstract}


\maketitle

The vertical segregation of particle sizes, also known as the Brazil nut problem \cite{RosatoEtal:1987}, has long been known but is very complex in its nature. A main reason for the complexity of the problem is that there exist different driving mechanisms for different experimental parameters, like, e.g., container shape and size, excitation acceleration and frequency, among others, and that these mechanisms are not mutually independent but can overlap and take place at the same time. One widespread driving mechanism is granular convection as shown by \citet{KnightEtal:1993, KnightEtal:1996}, who studied the convective motion of glass beads in long cylindrical Pyrex and Lucite containers by using dyed tracer particles as well as magnetic resonance imaging. With this mechanism, a collective motion of the medium can transport larger particles to the top, which come to rest there if the downward flow zone is too small to be entered by these particles. This effect has been studied experimentally \cite{KnightEtal:1993, KnightEtal:1996, DuranEtal:1994, VanelEtal:1997, GarcimartinEtal:2000, CookeEtal:1996, MoebiusEtal:2005, HejmadyEtal:2012, AhmadSmalley:1973, RodriguezNahmad:2006} and theoretically \cite{Rajchenbach:1991, Grossman:1997, RamirezEtal:2000, KlongboonjitCampbell:2008}, while the focus of the article at hand is on the extrapolation of the Brazil nut problem to reduced gravity conditions found on small Solar System bodies. The Brazil nut effect has been, for example, made responsible for observed surface structures on small asteroids \cite{AsphaugEtal:2001, MiyamotoEtal:2007}. Since many small bodies in the Solar System possess a granular surface \cite{GundlachBlum:2013} there are also technical aspects to this problem, like, e.g., the handling of materials on the Moon, material sampling on upcoming asteroid landing missions, or on larger scales even future asteroid mining (e.g., considering density segregation). For all of these problems, it is desirable to understand the scaling of the granular flow and, thus, the segregation timescale with the ambient acceleration $g_\mathrm{amb}$, which can be as small as $10^{-5}g_\mathrm{Earth}$ on a small asteroid.

\begin{figure}[t]
    \begin{center}
      \includegraphics[height=6.2cm]{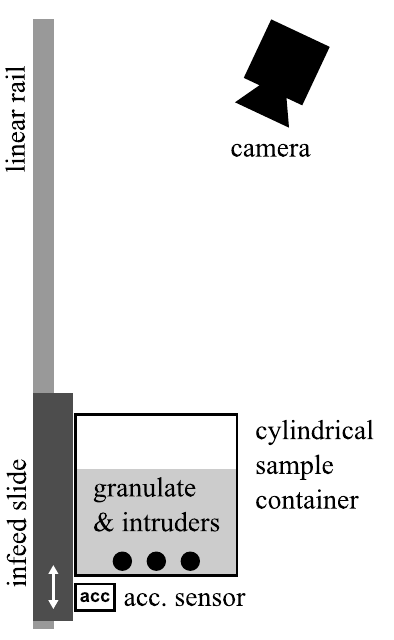}
			\includegraphics[trim=0cm 0cm 13.7cm 0cm, height=6.2cm]{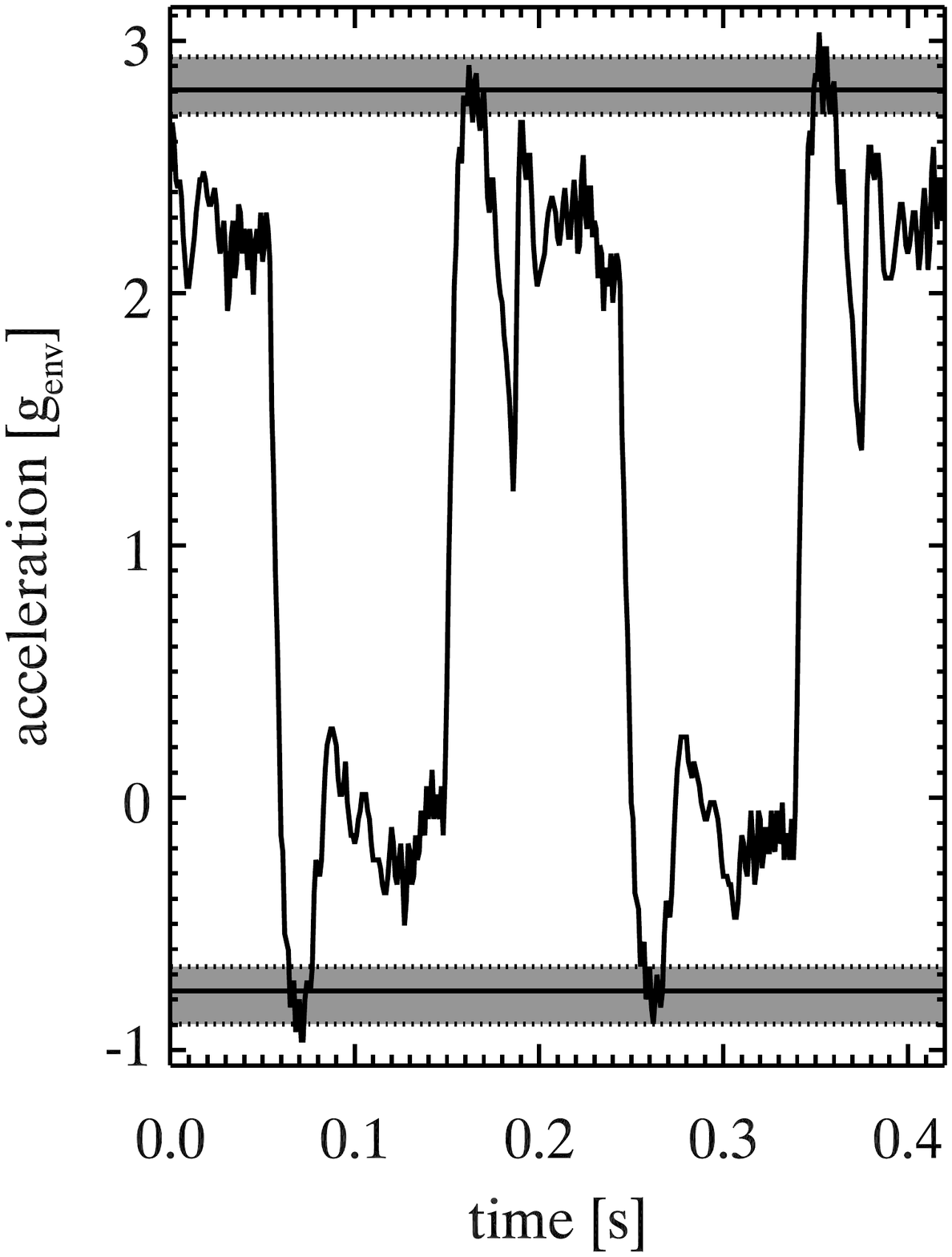}
      \caption{\label{fig:shaker}Left: Sketch of the experimental setup. The sample container is attached to an infeed slide, which can be moved vertically on a linear rail. Right: Acceleration profile of the infeed slide and the attached experiment container for an experiment with $g_\mathrm{amb}=g_\mathrm{Mars}$.}
    \end{center}
\end{figure}

We studied the Brazil nut effect of glass-bead samples in a transparent polycarbonate container of 110 mm inner diameter and 100 mm height. This container was firmly mounted on an infeed slide of a linear rail to vertically shake it in the direction of its cylindrical axis (Fig. \ref{fig:shaker}, left). The shaking profile in all experiments had an oscillation amplitude of $\pm10$ mm and was defined by an approximated square-wave function of the acceleration (see Fig. \ref{fig:shaker}, right). Depending on the individual experimental run, a pre-defined acceleration level was set to the control software of the linear stage and the acceleration was internally regulated. We additionally attached an external acceleration sensor to the infeed slide to measure the acceleration level and found a small overshoot before reaching a quasi-constant acceleration plateau at the pre-defined value (Fig. \ref{fig:shaker}, right). We will use the maximum acceleration amplitude of the periodic accelerometer signal as indicated by the gray-shaded areas for the excitation acceleration $a_\mathrm{exc}$. One should keep in mind that the vibration frequency was not kept constant with this setup, but changed from 7.0 to 11.1 Hz (for Earth-level gravitational acceleration), 5.1 to 5.7 Hz (Mars) and 3.4 to 3.8 Hz (Moon) for the the different experimental runs.

We used soda-lime glass beads of 1 mm diameter for the bulk material and 8 mm diameter for the intruders. The test chambers were initially filled to a height of a few millimeters with the small particles, then seven intruder beads were fixed in this granular bed in a hexagonal shape with one bead in the center, before the container was filled up with 1 mm beads to a nominal height of 60 mm in most cases. We do not expect an influence on the rise time due to multiple intruders, as confirmed by \citet{MoebiusEtal:2005}. The experiments were performed under normal air pressure and we therefore cannot exclude gas effects. However, due to the chosen particle size, mass density of the bulk material, and an equal density between bulk and intruder particles, we expect that the particle rise time is only slightly affected \cite{MoebiusEtal:2005}. The upper surface of the sample was observed with a digital camera with a precise internal timer to measure the rise time of the glass bead intruders, which we took with an accuracy of 1 s. A horizontal movement of the surface layer and some heaping was visible in the camera images. Moreover we observed that large beads, which appear at the surface, reemerge into the medium near the side walls and reappear after a time, which is always less than twice the rise time, so that we conclude to be in a convective regime.

Some experiments were performed on ground and, thus, under Earth gravity ($g_\mathrm{Earth} = 9.81$~\acc) and we measured the rise time (rise velocity) as a function of the fill height and the excitation acceleration, as will be presented below. The main focus of this work was, however, on low-gravity experiments, which were performed on board the A300 Zero-G aircraft flying a maneuver to achieve Martian ($g_\mathrm{Mars} = 3.71$ \acc) and Lunar gravity ($g_\mathrm{Moon} = 1.62$ \acc) levels. In these experiments, the excitation of the granular sample was started when the desired gravity level was attained and the excitation was stopped shortly before the transition to the following hyper-gravity maneuver. This yielded a time between 20 and 30 s per parabola and if the intruders did not appear in the first parabola, the experiment was continued in the subsequent parabola. The granular medium was perfectly in rest when the excitation was stopped so that the experimental runtime can simply be cumulated. In few cases, the excitation was only stopped within the hyper-gravity phase, but at the abrupt increase of the gravity level, the granular medium always came to a complete rest, which was clearly visible on the camera images. In these cases, we chose the time at which the granular medium stopped moving at the end of the experimental runtime of that particular parabola. In each of the four flights, in which 12 Mars- and 12 Moon-level parabolas were flown, we used several experiment containers, which were exchanged after all seven intruders had appeared.

\begin{figure}[t]
    \begin{center}
      \includegraphics[width=8.6cm]{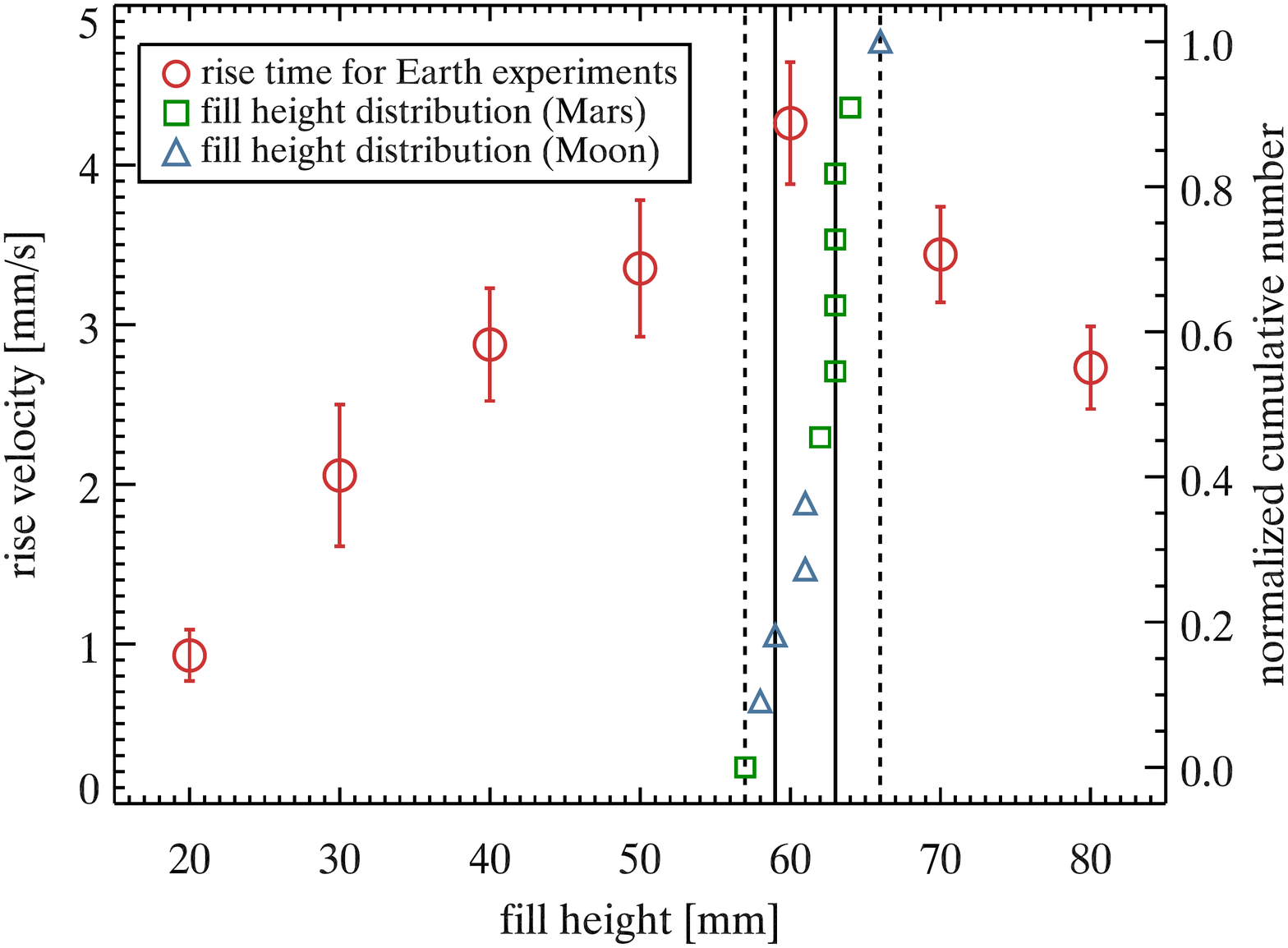}
      \caption{\label{fig:height_vs_velocity}The mean rise velocity as a function of the fill height of the experiment containers (red circles, left ordinate). Also plotted is a cumulative distribution of fill heights in the parabolic flight experiments (green squares and blue triangles, right ordinate), which shows that the range of fill heights (58 to 66 mm) was close to the maximum rise velocity.}
    \end{center}
\end{figure}

In a first set of experiments on Earth, we kept the amplitude of the excitation acceleration constant at a normalized value of $\Gamma=1.72$. Here, $\Gamma=a_\mathrm{exc}/g_\mathrm{amb}$ is the ratio of the excitation acceleration and the ambient gravitational acceleration (in this case with $g_\mathrm{amb}=g_\mathrm{Earth}$). We varied the fill height from 20 to 80 mm and placed the intruder beads always at the bottom of the container. The mean rise velocity (i.e., the fill height minus the intruder diameter, divided by the rise time) is shown in Fig. \ref{fig:height_vs_velocity} as a function of fill height. One experiment with seven intruders is represented by one red circle with the standard deviation shown by the error bars. We found a linear increase of the rise velocity for fill heights up to 60 mm. This is consistent with the observation of \citet{GarcimartinEtal:2000}, who found the flow velocity (downward flow near the wall) in their experiments to scale linearly with the number of layers at aspect ratios (container height divided by container diameter) from 0.18 to 0.45 but at higher excitation frequencies of 110 Hz. For larger fill heights, we found a decrease of the rise velocity, which has been studied earlier by \citeauthor{KnightEtal:1996} and \citeauthor{Grossman:1997} \cite{KnightEtal:1996, Grossman:1997}. These authors found an exponential decay of the rise velocity with increasing embedding depth for much larger aspect ratios than ours. For aspect ratios less than unity, they found that the rise velocity decreases with increasing fill height (but with an unknown relation), so that our results are also consistent with these findings. We are not aware of any publication showing an abrupt change in the rise velocity as presented in Fig. \ref{fig:height_vs_velocity}. The fill height for the Earth-gravity experiments was precise to within one millimeter, but due to different filling procedures in the low-gravity parabolic flight experiments, the latter show a small variation. Thus, we also plotted in Fig. \ref{fig:height_vs_velocity} a cumulative distribution of fill heights of the parabolic-flight experiments, which are between 58 and 66 mm and, thus, in a regime with maximal rise velocity.

\begin{figure}[t]
    \begin{center}
      \includegraphics[width=8.6cm]{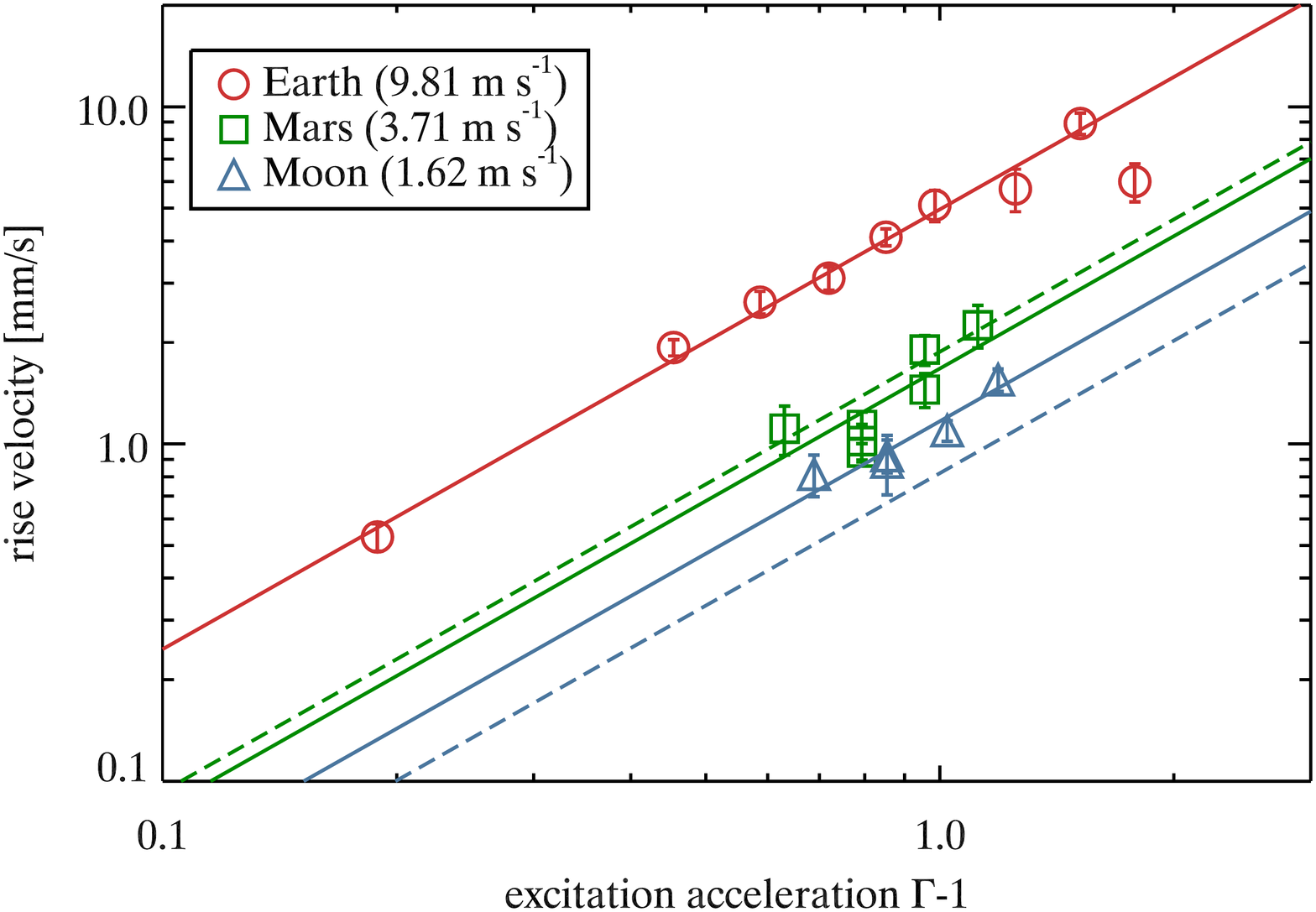}
      \caption{\label{fig:acc_vs_time}Rise velocity for experiments at terrestrial (red circles), Martian (green squares), and Lunar gravity values (blue triangles) as a function of the excitation acceleration, here represented by $\Gamma-1$. The error bars show the standard deviation in one experimental run with seven intruders. The solid and dashed lines represent power laws with an exponent of 1.3 as described in the text.}
    \end{center}
\end{figure}

Figure \ref{fig:acc_vs_time} shows the rise velocity of the terrestrial and parabolic-flight experiments, in which the fill height was either 60 mm (Earth) or 58 to 66 mm (Mars and Moon) as a function of the excitation acceleration $\Gamma-1$. For the terrestrial experiments, where $g_\mathrm{amb}=g_\mathrm{Earth}$, we find that the data are well reproduced by a power law of the form
\begin{equation}
	v_\mathrm{rise} \propto (\Gamma-1)^{1.30}\ , \label{eq:t_rise}
\end{equation}
thus the choice of $\Gamma-1$ as the horizontal axis. We did not use the data point with the highest $\Gamma-1$ value for the fit as it clearly deviates from the other data for an unknown reason. A vanishing velocity at $\Gamma \lesssim 1$ is consistent with most earlier works \cite{KnightEtal:1996, DuranEtal:1994, VanelEtal:1997, GarcimartinEtal:2000}. To compare our results to the data of \citet{KnightEtal:1996}, we translated our rise times into tap numbers, thus multiplied it with the oscillation frequency. Considering their data for 1 mm glass beads in Figs. 3a and 4a and their Eq. 1, we can see a rough quantitative agreement for an initial depth of $z=60$ mm, but we can get a perfect match (slope and absolute value) for $z=30$ mm. A possible reason to choose an initial depth smaller than our fill height is their smaller container diameter, which is smaller than ours by a factor of two. We could not find an agreement with other publications, which were however either sparsely described \cite{AhmadSmalley:1973} or used a 2D setup \cite{CookeEtal:1996, HejmadyEtal:2012}. Available data for the convection velocity at the walls, which show an exponent in Eq. \ref{eq:t_rise} between 1 and 2 \cite{GarcimartinEtal:2000, RodriguezNahmad:2006}, cannot directly be compared to our data as \citet{HejmadyEtal:2012} showed that the exponent for the near-wall velocity is close to unity, while the rise velocity (as derived from their rise times) possesses a much steeper dependence on excitation acceleration.

The excitation acceleration and the rise velocity in the parabolic-flight experiments was measured the same way as for the terrestrial experiments. However, since the ambient gravitational acceleration (i.e., Lunar or Martian gravity) is not really constant (because the aircraft encounters turbulence), we have to apply a correction to the rise-velocity data. Even though the mean acceleration over a full parabola comes very close to the nominal Lunar or Martian gravity level, an oscillation of typically $\pm0.03g_\mathrm{Earth}$ around this value has a non-negligible impact on the rise velocity. If the aircraft acceleration increases (decreases) over the nominal value, the relative excitation $\Gamma$ gets smaller (higher). The corrected rise velocities \cite{PRL_note_1} are shown as green squares (Martian experiments) and blue triangles (Lunar experiments) in Fig. \ref{fig:acc_vs_time}, again representing mean values for one container with seven intruders. The scatter of these data is slightly larger than for the terrestrial experiments -- most likely still due to variations in the ambient gravity level -- but a clear trend of increasing rise velocity with increasing excitation level is evident. A power-law fit to the two low-gravity data sets yields exponents of 1.47 (Mars) and 1.17 (Moon), respectively, which are close enough to the terrestrial value of 1.30, which is based on a much wider range of excitation accelerations. The green and blue solid lines in Fig. \ref{fig:acc_vs_time} represent power laws with a slope of 1.30. The dashed lines represent the expected velocities for Martian and Lunar gravity levels if the scaling with gravity is linear. The Martian data are consistent with this linear extrapolation, while the Lunar data significantly deviate from the linear trend.

\begin{figure}[t]
    \begin{center}
      \includegraphics[width=8.6cm]{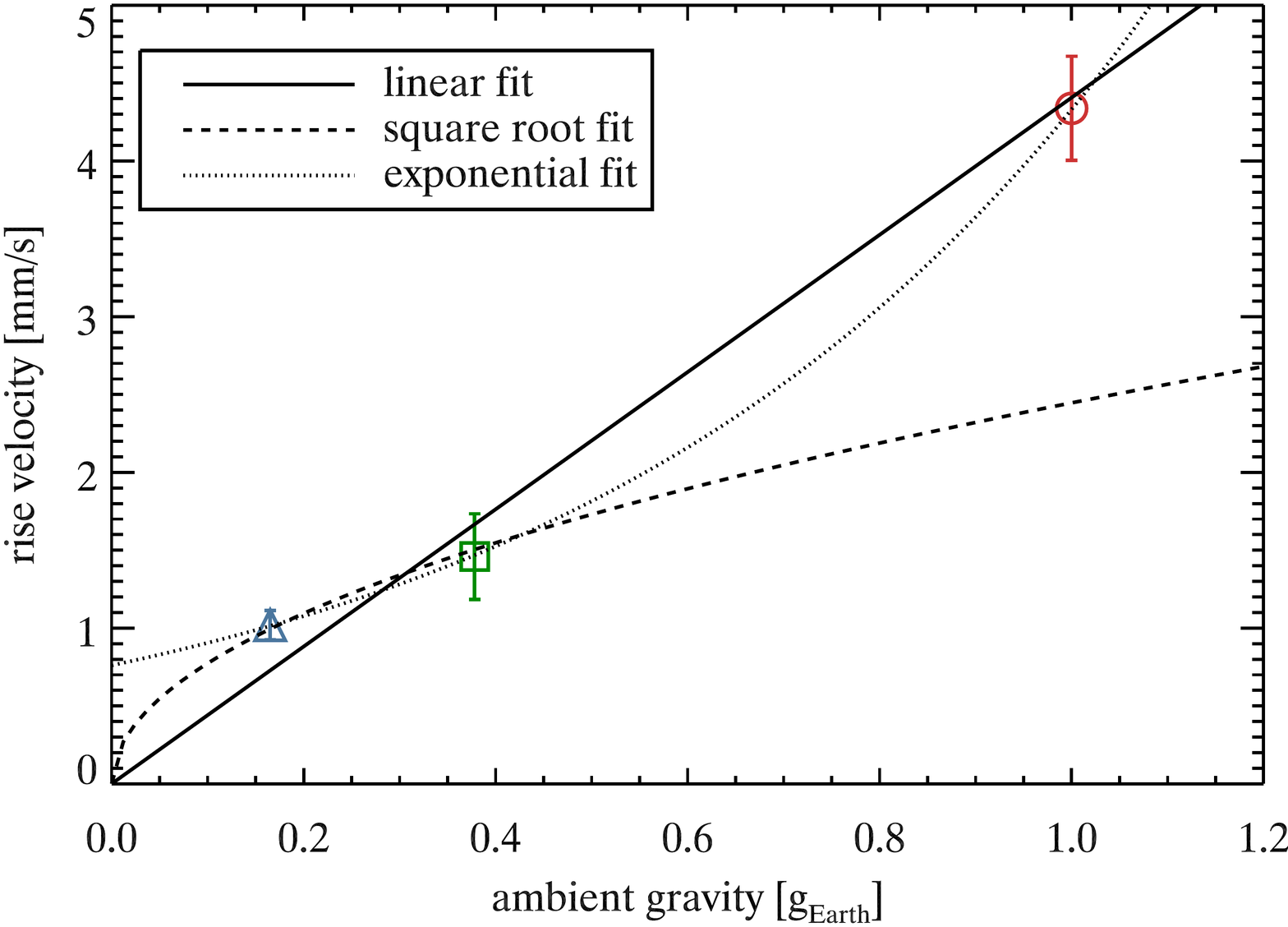}
      \caption{\label{fig:grav_vs_time}Rise velocity for $\Gamma=1.9$ as a function of the ambient gravity level $g_\mathrm{amb}$ for the three gravity regimes studied.}
    \end{center}
\end{figure}

To derive the relation between the rise velocity and ambient gravity level, we chose a reference point of $\Gamma=1.9$ for the normalized excitation acceleration (roughly the median acceleration of all experiments). For this normalization, each data point was shifted according to Eq. \ref{eq:t_rise} and the highest $\Gamma-1$ value for the Earth experiments, which was also not used for the fit, was again neglected. The resulting mean rise velocity of all experiments in one gravity regime is shown as one data point in Fig. \ref{fig:grav_vs_time} (same symbols as in Fig. \ref{fig:acc_vs_time}) for the three ambient gravitational accelerations. The error bars represent standard deviations of the mean values. One can see that the rise time is roughly in a linear relation to the ambient gravity level, as indicated by the solid line. The deviation of the rise time for the Lunar experiments is, however, statistically significant so that we cannot make a firm statement on the functional behavior. A perfect fit can actually be achieved with an exponential function, shown by the dotted line, with the drawback that it gives an implausible non-zero rise velocity for a vanishing ambient acceleration, if gravity is the driving force for the convective motion. A linear relation is more plausible, but at low gravity levels we expect the highest deviations, because (1) cohesion forces between the particles become important, (2) friction becomes less important due to smaller normal forces, and (3) the role of collisions changes due to the velocity dependence of the coefficient of restitution. Without a numerical model it is, however, difficult to predict whether we expect the convection to be enhanced or damped at low gravity levels.

A linear relation between the convective flux (thus, the rise velocity) and gravity was predicted in the model of \citet{Rajchenbach:1991}, which is based on the dilatancy of the granular medium and the gradients in density and mobility. Other models based on thermal convection \cite{RamirezEtal:2000} or wall friction \cite{KlongboonjitCampbell:2008} describe a 'slight' or square root dependence on gravity, respectively. With our presented gravity dependence, we provide a new parameter to verify (or falsify) granular-convection models and we can already rule out any model with a vanishing or weak dependence between convective velocity and gravitational level. We should, however, be careful with a too strong statement, because we can provide only three data points and are therefore unable to discriminate two superimposing effects (e.g., convection plus percolation) with a different gravity dependence. In fact, the rise velocity for the Lunar and Martian experiments are in a perfect square-root dependence on velocity (dashed line in Fig. \ref{fig:grav_vs_time}), and an additional superimposed effect could explain the faster rise velocity for the terrestrial experiments.

The strong deviation between our linear fit and the rise velocity at the Lunar gravity level inhibits an extrapolation to much lower gravity levels. A promising approach to improve our knowledge on the gravity-level dependence of the rise velocity would be experiments under the enhanced acceleration conditions of a centrifuge, which could establish a reliable scaling of the convection velocity over orders of magnitude. In the meantime, physical and numerical models for granular convection should be reconsidered to explain a (roughly) linear dependence on the ambient gravity.

We thank Pilipp Reiss, Ralf Pursche, and Martin Rott for sharing their parabolic flight with us and for helping us with the preparation and performance of the experiments. The parabolic flight was kindly provided by ESA and the experimental hardware was funded by DLR through grant 50WM1236. C.G. is grateful to the Japan Society for the Promotion of Science for his funding.

\bibliography{literature,prl_notes}

\begin{thebibliography}{19}
\bibitem[{PRL()}]{PRL_note_1}
\bibinfo{note}{The rise time, which determines the rise velocity, was
  calculated by adding up the time intervals $\Delta t$ of the aircraft's
  acceleration sensor during the time of excitation and individually weight
  them with $[a_\mathrm{nom}/a_\mathrm{real}(t)]^{-1.3}$. Here,
  $a_\mathrm{nom}=a_\mathrm{exc}/g_\mathrm{amb}-1 = \Gamma-1$ is the desired
  excitation acceleration and $a_\mathrm{real}(t) = a_\mathrm{exc}/g(t)-1$ is
  the actual excitation relative to the aircraft gravity $g(t)$ at time $t$.
  The slope of -1.3 is the inverse slope of Eq. \ref{eq:t_rise} and was
  confirmed in an independent analysis of the rise time and excitation
  acceleration. This correction changes the rise times and thus velocities by
  approximately 10\%.}
\end{thebibliography}

\end{document}